\documentstyle[10pt,epsf,epsfig,dp_delphititle,oldlfont,amsmath]{dp_delphi}
\makeindex
\def\DpPaperGroup{EP}
\def\DpPaperRef{2001-006}
\def\DpDate{8 January-2001}
\def\DpAuthors{DELPHI Collaboration}
\def\DpSubmit{(Accepted by Phys.Lett.B)}
\def\DpTitle{{\bf Measurement of Trilinear Gauge Boson Couplings $WWV$,
($V \equiv Z, \gamma$) in \ee\ Collisions at 189~GeV
     }}
\def\DpComment{}
\def\DpEMail{}

\newcommand{\Dgz}{\mbox{$\Delta g_1^Z$}}

\newcommand{\kg}{\mbox{$\kappa_\gamma$}}
\newcommand{\Dkg}{\mbox{$\Delta\kappa_\gamma$}}

\newcommand{\lm}{\mbox{$\lambda_\gamma$}}
\newcommand{\lz}{\mbox{$\lambda_Z$}}

\newcommand{\cosw}{\mbox{$\cos \theta_W$}}
\newcommand{\cosl}{\mbox{$\cos \theta_\ell$}}

\newcommand{\ra}{\mbox{$\rightarrow$}}
\newcommand{\ee}{\mbox{$e^{+}e^{-}$}}
\newcommand{\jjlv} {\mbox{$ j j \ell \nu $}}
\newcommand{\jjjj} {\mbox{$ j j j j $}}

\newcommand{\jjX} {\mbox{$ j j X $}}
\newcommand{\lX} {\mbox{$ \ell X $}}
\newcommand{\gX} {\mbox{$ \gamma X $}}

\newcommand{\Wev}{\mbox{${W e} \nu $}}
\newcommand{\vvg}{\mbox{$ \nu \nu \gamma $}}
\newcommand{\Wp} {\mbox{$W^+ $}}
\newcommand{\Wm} {\mbox{$W^- $}}
\newcommand{\WpWm} {\mbox{$W^+ W^- $}}

\newcommand{\eeWW}{\mbox{\ee \ra \, \WpWm}}
\newcommand{\eeWev}{\mbox{\ee \ra \, \Wev}}
\newcommand{\eevvg}{\mbox{\ee \ra \, \vvg}}

\newcommand{\aWphi}{\mbox{$\alpha_{W\phi} $}}
\newcommand{\aBphi}{\mbox{$\alpha_{B\phi} $}}
\newcommand{\aW}{\mbox{$\alpha_{W} $}}

\newcommand{\dgr}{\mbox{$^\circ$}}

\def\pl#1#2#3{Phys.\ Lett.~{\bf B#1} (#3) #2}

\def\np#1#2#3{Nucl.\ Phys.~{\bf B#1} (#3) #2}
\def\zp#1#2#3{Z.\ Phys.~{\bf C#1} (#3) #2}
\def\nim#1#2#3{Nucl.\ Inst.\ Meth.~{\bf A#1} (#3) #2}
\def\epj#1#2#3{E.\ Phys.\ J.~{\bf C#1} (#3) #2}

\newcommand{\bq}{\begin{equation}}
\newcommand{\eq}{\end{equation}}
\newcommand{\ba}{\begin{eqnarray}}
\newcommand{\ea}{\end{eqnarray}}

\begin{document}
\makeatletter
\newcount\@tempcntc
\def\@citex[#1]#2{\if@filesw\immediate\write\@auxout{\string\citation{#2}}\fi
  \@tempcnta\z@\@tempcntb\m@ne\def\@citea{}\@cite{\@for\@citeb:=#2\do
    {\@ifundefined
       {b@\@citeb}{\@citeo\@tempcntb\m@ne\@citea\def\@citea{,}{\bf ?}\@warning
       {Citation `\@citeb' on page \thepage \space undefined}}%
    {\setbox\z@\hbox{\global\@tempcntc0\csname b@\@citeb\endcsname\relax}%
     \ifnum\@tempcntc=\z@ \@citeo\@tempcntb\m@ne
       \@citea\def\@citea{,}\hbox{\csname b@\@citeb\endcsname}%
     \else
      \advance\@tempcntb\@ne
      \ifnum\@tempcntb=\@tempcntc
      \else\advance\@tempcntb\m@ne\@citeo
      \@tempcnta\@tempcntc\@tempcntb\@tempcntc\fi\fi}}\@citeo}{#1}}
\def\@citeo{\ifnum\@tempcnta>\@tempcntb\else\@citea\def\@citea{,}%
  \ifnum\@tempcnta=\@tempcntb\the\@tempcnta\else
   {\advance\@tempcnta\@ne\ifnum\@tempcnta=\@tempcntb \else \def\@citea{--}\fi
    \advance\@tempcnta\m@ne\the\@tempcnta\@citea\the\@tempcntb}\fi\fi}
 
\makeatother
\begin{titlepage}
\pagenumbering{roman}
\CERNpreprint{\DpPaperGroup}{\DpPaperRef} 
\date{{\small\DpDate}} 
\title{\DpTitle} 
\address{\DpAuthors} 
\begin{shortabs} 
\noindent
%
Measurements of the trilinear gauge boson couplings $WW\gamma$ and
$WWZ$ are presented using the data taken by DELPHI in 1998 at a 
centre-of-mass energy of 189~GeV
and combined with DELPHI data at 183~GeV. Values are
determined for \Dgz\ and \Dkg, the differences of the $WWZ$ charge coupling and
of the $WW\gamma$ dipole coupling from their Standard Model values, and for
\lm, the $WW\gamma$ quadrupole coupling. 
A measurement of the magnetic dipole and electric quadrupole moment of the $W$
is extracted from the results for \Dkg{} and \lm. 
The study uses data from the final
states \jjlv, \jjjj, \lX, \jjX\ and \gX, where $j$ represents a quark
jet, $\ell$ an identified lepton and $X$ missing four-momentum.  The
observations are consistent with the predictions of the Standard Model.

\end{shortabs}
\vfill
\begin{center}
\DpSubmit \ \\ 
\DpComment \ \\
\DpEMail \ \\
\end{center}
\vfill
\clearpage
\headsep 10.0pt
\addtolength{\textheight}{10mm}
\addtolength{\footskip}{-5mm}
\begingroup
%
\newcommand{\DpName}[2]{\hbox{#1$^{\ref{#2}}$},\hfill}
\newcommand{\DpNameTwo}[3]{\hbox{#1$^{\ref{#2},\ref{#3}}$},\hfill}
\newcommand{\DpNameThree}[4]{\hbox{#1$^{\ref{#2},\ref{#3},\ref{#4}}$},\hfill}
\newskip\Bigfill \Bigfill = 0pt plus 1000fill
\newcommand{\DpNameLast}[2]{\hbox{#1$^{\ref{#2}}$}\hspace{\Bigfill}}
%
\footnotesize
\noindent
\DpName{P.Abreu}{LIP}
\DpName{W.Adam}{VIENNA}
\DpName{T.Adye}{RAL}
\DpName{P.Adzic}{DEMOKRITOS}
\DpName{Z.Albrecht}{KARLSRUHE}
\DpName{T.Alderweireld}{AIM}
\DpName{G.D.Alekseev}{JINR}
\DpName{R.Alemany}{CERN}
\DpName{T.Allmendinger}{KARLSRUHE}
\DpName{P.P.Allport}{LIVERPOOL}
% ...
\normalsize
\endgroup
\titlefoot{Department of Physics and Astronomy, Iowa State
     University, Ames IA 50011-3160, USA
    \label{AMES}}
\titlefoot{Physics Department, Univ. Instelling Antwerpen,
     Universiteitsplein 1, B-2610 Antwerpen, Belgium \\
     \indent~~and IIHE, ULB-VUB,
     Pleinlaan 2, B-1050 Brussels, Belgium \\
     \indent~~and Facult\'e des Sciences,
     Univ. de l'Etat Mons, Av. Maistriau 19, B-7000 Mons, Belgium
    \label{AIM}}
\titlefoot{Physics Laboratory, University of Athens, Solonos Str.
     104, GR-10680 Athens, Greece
    \label{ATHENS}}
\titlefoot{Department of Physics, University of Bergen,
     All\'egaten 55, NO-5007 Bergen, Norway
    \label{BERGEN}}
\titlefoot{Dipartimento di Fisica, Universit\`a di Bologna and INFN,
     Via Irnerio 46, IT-40126 Bologna, Italy
    \label{BOLOGNA}}
\titlefoot{Centro Brasileiro de Pesquisas F\'{\i}sicas, rua Xavier Sigaud 150,
     BR-22290 Rio de Janeiro, Brazil \\
     \indent~~and Depto. de F\'{\i}sica, Pont. Univ. Cat\'olica,
     C.P. 38071 BR-22453 Rio de Janeiro, Brazil \\
     \indent~~and Inst. de F\'{\i}sica, Univ. Estadual do Rio de Janeiro,
     rua S\~{a}o Francisco Xavier 524, Rio de Janeiro, Brazil
    \label{BRASIL}}
\titlefoot{Comenius University, Faculty of Mathematics and Physics,
     Mlynska Dolina, SK-84215 Bratislava, Slovakia
    \label{BRATISLAVA}}
\titlefoot{Coll\`ege de France, Lab. de Physique Corpusculaire, IN2P3-CNRS,
     FR-75231 Paris Cedex 05, France
    \label{CDF}}
\titlefoot{CERN, CH-1211 Geneva 23, Switzerland
    \label{CERN}}
\titlefoot{Institut de Recherches Subatomiques, IN2P3 - CNRS/ULP - BP20,
     FR-67037 Strasbourg Cedex, France
    \label{CRN}}
\titlefoot{Now at DESY-Zeuthen, Platanenallee 6, D-15735 Zeuthen, Germany
    \label{DESY}}
\titlefoot{Institute of Nuclear Physics, N.C.S.R. Demokritos,
     P.O. Box 60228, GR-15310 Athens, Greece
    \label{DEMOKRITOS}}
\titlefoot{FZU, Inst. of Phys. of the C.A.S. High Energy Physics Division,
     Na Slovance 2, CZ-180 40, Praha 8, Czech Republic
    \label{FZU}}
\titlefoot{Dipartimento di Fisica, Universit\`a di Genova and INFN,
     Via Dodecaneso 33, IT-16146 Genova, Italy
    \label{GENOVA}}
\titlefoot{Institut des Sciences Nucl\'eaires, IN2P3-CNRS, Universit\'e
     de Grenoble 1, FR-38026 Grenoble Cedex, France
    \label{GRENOBLE}}
\titlefoot{Helsinki Institute of Physics, HIP,
     P.O. Box 9, FI-00014 Helsinki, Finland
    \label{HELSINKI}}
\titlefoot{Joint Institute for Nuclear Research, Dubna, Head Post
     Office, P.O. Box 79, RU-101 000 Moscow, Russian Federation
    \label{JINR}}
\titlefoot{Institut f\"ur Experimentelle Kernphysik,
     Universit\"at Karlsruhe, Postfach 6980, DE-76128 Karlsruhe,
     Germany
    \label{KARLSRUHE}}
\titlefoot{Institute of Nuclear Physics and University of Mining and Metalurgy,
     Ul. Kawiory 26a, PL-30055 Krakow, Poland
    \label{KRAKOW}}
\titlefoot{Universit\'e de Paris-Sud, Lab. de l'Acc\'el\'erateur
     Lin\'eaire, IN2P3-CNRS, B\^{a}t. 200, FR-91405 Orsay Cedex, France
    \label{LAL}}
\titlefoot{School of Physics and Chemistry, University of Lancaster,
     Lancaster LA1 4YB, UK
    \label{LANCASTER}}
\titlefoot{LIP, IST, FCUL - Av. Elias Garcia, 14-$1^{o}$,
     PT-1000 Lisboa Codex, Portugal
    \label{LIP}}
\titlefoot{Department of Physics, University of Liverpool, P.O.
     Box 147, Liverpool L69 3BX, UK
    \label{LIVERPOOL}}
\titlefoot{LPNHE, IN2P3-CNRS, Univ.~Paris VI et VII, Tour 33 (RdC),
     4 place Jussieu, FR-75252 Paris Cedex 05, France
    \label{LPNHE}}
\titlefoot{Department of Physics, University of Lund,
     S\"olvegatan 14, SE-223 63 Lund, Sweden
    \label{LUND}}
\titlefoot{Universit\'e Claude Bernard de Lyon, IPNL, IN2P3-CNRS,
     FR-69622 Villeurbanne Cedex, France
    \label{LYON}}
\titlefoot{Univ. d'Aix - Marseille II - CPP, IN2P3-CNRS,
     FR-13288 Marseille Cedex 09, France
    \label{MARSEILLE}}
\titlefoot{Dipartimento di Fisica, Universit\`a di Milano and INFN-MILANO,
     Via Celoria 16, IT-20133 Milan, Italy
    \label{MILANO}}
\titlefoot{Dipartimento di Fisica, Univ. di Milano-Bicocca and
     INFN-MILANO, Piazza delle Scienze 2, IT-20126 Milan, Italy
    \label{MILANO2}}
\titlefoot{IPNP of MFF, Charles Univ., Areal MFF,
     V Holesovickach 2, CZ-180 00, Praha 8, Czech Republic
    \label{NC}}
\titlefoot{NIKHEF, Postbus 41882, NL-1009 DB
     Amsterdam, The Netherlands
    \label{NIKHEF}}
\titlefoot{National Technical University, Physics Department,
     Zografou Campus, GR-15773 Athens, Greece
    \label{NTU-ATHENS}}
\titlefoot{Physics Department, University of Oslo, Blindern,
     NO-1000 Oslo 3, Norway
    \label{OSLO}}
\titlefoot{Dpto. Fisica, Univ. Oviedo, Avda. Calvo Sotelo
     s/n, ES-33007 Oviedo, Spain
    \label{OVIEDO}}
\titlefoot{Department of Physics, University of Oxford,
     Keble Road, Oxford OX1 3RH, UK
    \label{OXFORD}}
\titlefoot{Dipartimento di Fisica, Universit\`a di Padova and
     INFN, Via Marzolo 8, IT-35131 Padua, Italy
    \label{PADOVA}}
\titlefoot{Rutherford Appleton Laboratory, Chilton, Didcot
     OX11 OQX, UK
    \label{RAL}}
\titlefoot{Dipartimento di Fisica, Universit\`a di Roma II and
     INFN, Tor Vergata, IT-00173 Rome, Italy
    \label{ROMA2}}
\titlefoot{Dipartimento di Fisica, Universit\`a di Roma III and
     INFN, Via della Vasca Navale 84, IT-00146 Rome, Italy
    \label{ROMA3}}
\titlefoot{DAPNIA/Service de Physique des Particules,
     CEA-Saclay, FR-91191 Gif-sur-Yvette Cedex, France
    \label{SACLAY}}
\titlefoot{Instituto de Fisica de Cantabria (CSIC-UC), Avda.
     los Castros s/n, ES-39006 Santander, Spain
    \label{SANTANDER}}
\titlefoot{Dipartimento di Fisica, Universit\`a degli Studi di Roma
     La Sapienza, Piazzale Aldo Moro 2, IT-00185 Rome, Italy
    \label{SAPIENZA}}
\titlefoot{Inst. for High Energy Physics, Serpukov
     P.O. Box 35, Protvino, (Moscow Region), Russian Federation
    \label{SERPUKHOV}}
\titlefoot{J. Stefan Institute, Jamova 39, SI-1000 Ljubljana, Slovenia
     and Laboratory for Astroparticle Physics,\\
     \indent~~Nova Gorica Polytechnic, Kostanjeviska 16a, SI-5000 Nova Gorica, Slovenia, \\
     \indent~~and Department of Physics, University of Ljubljana,
     SI-1000 Ljubljana, Slovenia
    \label{SLOVENIJA}}
\titlefoot{Fysikum, Stockholm University,
     Box 6730, SE-113 85 Stockholm, Sweden
    \label{STOCKHOLM}}
\titlefoot{Dipartimento di Fisica Sperimentale, Universit\`a di
     Torino and INFN, Via P. Giuria 1, IT-10125 Turin, Italy
    \label{TORINO}}
\titlefoot{Dipartimento di Fisica, Universit\`a di Trieste and
     INFN, Via A. Valerio 2, IT-34127 Trieste, Italy \\
     \indent~~and Istituto di Fisica, Universit\`a di Udine,
     IT-33100 Udine, Italy
    \label{TU}}
\titlefoot{Univ. Federal do Rio de Janeiro, C.P. 68528
     Cidade Univ., Ilha do Fund\~ao
     BR-21945-970 Rio de Janeiro, Brazil
    \label{UFRJ}}
\titlefoot{Department of Radiation Sciences, University of
     Uppsala, P.O. Box 535, SE-751 21 Uppsala, Sweden
    \label{UPPSALA}}
\titlefoot{IFIC, Valencia-CSIC, and D.F.A.M.N., U. de Valencia,
     Avda. Dr. Moliner 50, ES-46100 Burjassot (Valencia), Spain
    \label{VALENCIA}}
\titlefoot{Institut f\"ur Hochenergiephysik, \"Osterr. Akad.
     d. Wissensch., Nikolsdorfergasse 18, AT-1050 Vienna, Austria
    \label{VIENNA}}
\titlefoot{Inst. Nuclear Studies and University of Warsaw, Ul.
     Hoza 69, PL-00681 Warsaw, Poland
    \label{WARSZAWA}}
\titlefoot{Fachbereich Physik, University of Wuppertal, Postfach
     100 127, DE-42097 Wuppertal, Germany
    \label{WUPPERTAL}}
\addtolength{\textheight}{-10mm}
\addtolength{\footskip}{5mm}
\clearpage
\headsep 30.0pt
\end{titlepage}
%
\pagenumbering{arabic} 
\setcounter{footnote}{0} %
\large
%
%
\section{Introduction}

The properties of the final state in the reactions \eeWW, \Wev\ and \vvg\ are
sensitive to trilinear gauge boson couplings~\cite{DELPHI172,DELPHI183}.
This study uses data from the final states
\jjlv, \jjjj, \lX, \jjX\ and \gX\ (where $j$ represents a quark jet,
$\ell$ an identified lepton and $X$ missing four-momentum) taken by the DELPHI
detector at LEP in 1998 at a centre-of-mass energy of 189~GeV. The data are
used to determine values of three coupling parameters at the $WWV$ vertex (with
$V \equiv \gamma,Z$): \Dgz, the difference between the value of the overall
$WWZ$ coupling strength and its Standard Model prediction; \Dkg, the difference
between the value of the  dipole coupling, \kg, and its Standard Model value;
and \lm, the $WW\gamma$ quadrupole coupling parameter~\cite{Hagiwara}. 

In the evaluation of the couplings, a model has been assumed~\cite{YB} in
which contributions to the effective $WWV$ Lagrangian from operators describing
possible new physics beyond the Standard Model are restricted to those which are
$CP$-conserving, are of lowest dimension ($\leq 6$), satisfy $SU(2) \times U(1)$
invariance, and have not been excluded by previous measurements. 
This leads to
possible contributions from three operators, ${\cal L}_{W\phi}$, ${\cal
L}_{B\phi}$ and ${\cal L}_W$, and hence to relations between the permitted
values of the $WW\gamma$ and $WWZ$  couplings: $\Delta\kappa_Z =  \Dgz -
\frac{s_w^2}{c_w^2}\Dkg$ and $\lz = \lm$, where $s_w$ and $c_w$ are the sine and
cosine of the electroweak mixing angle. The parameters we determine are related
to possible contributions \aWphi, \aBphi\ and \aW\ from the three operators
given above by: $\Dgz =  \aWphi / c_w^2$, $\Dkg = \aWphi + \aBphi$, and $\lm =
\aW$. 

\begin{figure}[htb]
\centerline{\epsfig {file=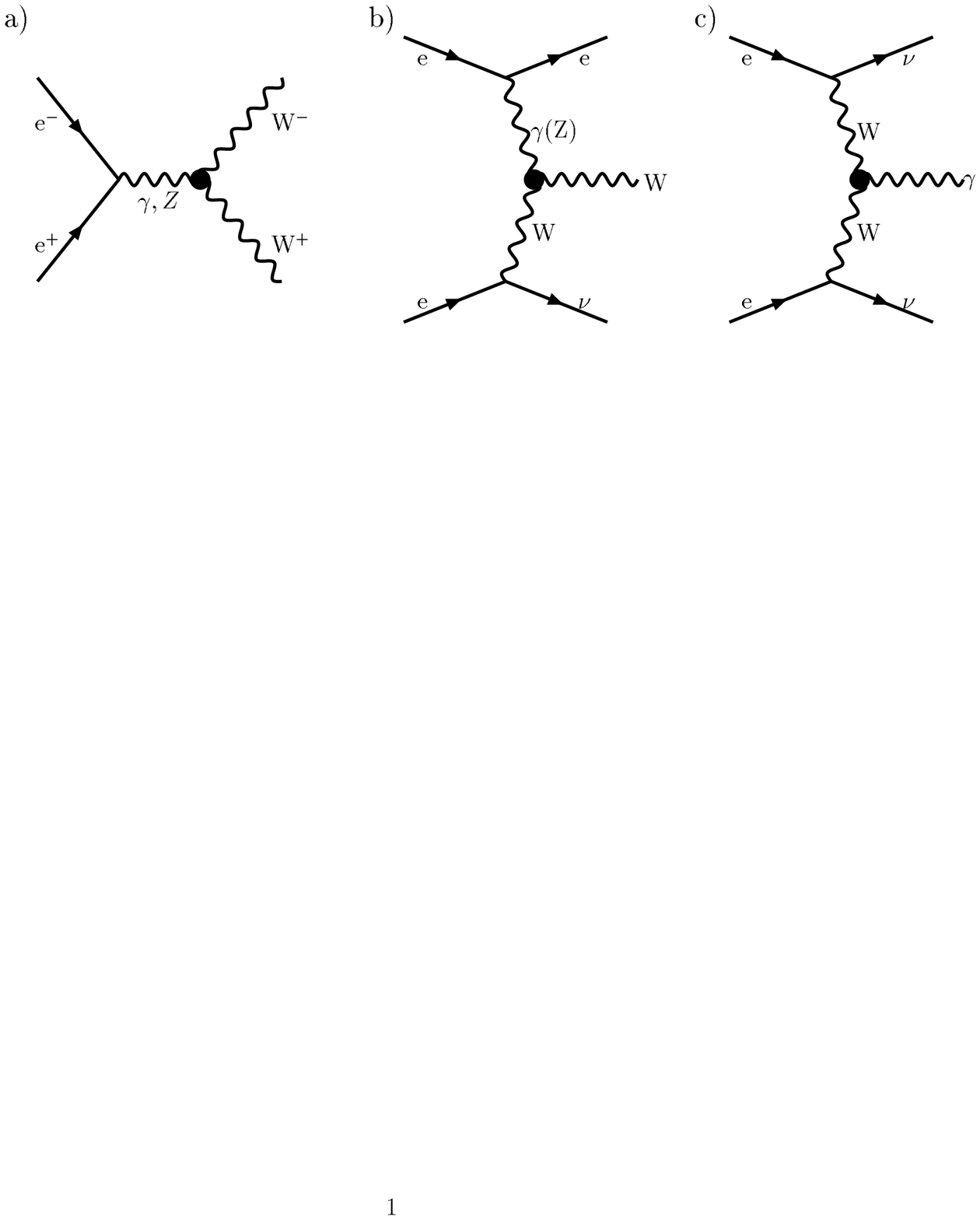,width=\textwidth,clip=} }
\caption[]{
Diagrams with trilinear gauge boson couplings contributing to the processes
studied in this paper: a) \eeWW, b) \eeWev, c) \eevvg.
}
\label{fig:Fdiags} 
\end{figure}

The $WWV$ coupling arises in $WW$ production through the diagrams involving
$s$-channel exchange of $Z$ or $\gamma$, shown in figure~\ref{fig:Fdiags}a. We
study this reaction  in the final states \jjlv, where one $W$ decays into
quarks and the other into leptons, and \jjjj, where both $W$s decay into quarks.

In single $W$ production, the dominant amplitude involving a trilinear gauge
coupling arises from the radiation of a virtual photon from the incident
electron or positron, interacting with a virtual $W$ radiated from the other
incident particle (figure~\ref{fig:Fdiags}b). This process, involving
a $WW\gamma$ coupling, contributes significantly in the kinematic region where
a final state electron or positron is emitted at small angle to the beam and is
thus likely to remain undetected in the beam pipe. The decay modes of the $W$
give rise to two final states: that with two jets and missing energy (\jjX), and
that containing only a single visible lepton coming from the interaction
point and no other track in the detector (\lX).   

The trilinear $WW\gamma$ vertex also occurs in the reaction \eevvg\ in the
diagram in which the incoming electron and positron each radiate a virtual $W$
at an $e\nu W$ vertex and these two fuse to produce an outgoing photon
(figure~\ref{fig:Fdiags}c). In this process, which leads to a final state, \gX,
consisting of a single detected photon, the $WW\gamma$ coupling is studied
completely independently of the $WWZ$ coupling, as no $WWZ$ vertex is involved.

The next section of this paper describes the selection of events from the data
and the simulation of the various channels involved in the analysis.
Section~\ref{sec:methods} describes the methods used in the  determination of
coupling parameters. In section~\ref{sec:results} the results from different
channels are presented and combined with previously published DELPHI
results~\cite{DELPHI183} to give overall values for the coupling
parameters. A summary is given in section~\ref{sec:Conclusions}.

\section{Event simulation and selection}
\label{sec:events}

In 1998 DELPHI recorded a total integrated luminosity of 155~pb$^{-1}$  at an
average centre-of-mass energy of $188.63 \pm 0.04$ GeV. We characterise here
the main features of 
the selection of events in the final state topologies \jjlv, \jjjj, \lX,
\jjX\ and \gX, defined in the previous section. A detailed description of the
DELPHI detector may be found in~\cite{detector}, which includes descriptions of
the main components of the detector used in this study, namely, the trigger
system, the luminosity monitor, the tracking system in the barrel and forward
regions, the muon detectors, the electromagnetic calorimeters and the
hermeticity counters. The definition of the criteria imposed for track
selection and lepton identification and a description of the luminosity
measurement are given in~\cite{W97}. 

\vspace{0.5cm}
\noindent {\bf Event simulation:}
\vspace{0.2cm}

\noindent
Various Monte Carlo models were used in the calculation of cross-sections  as a
function of coupling parameters in the different final states analysed. In the
study of the \jjlv\ and \jjjj\ channels, the four-fermion generators
EXCALIBUR~\cite{EXCAL} and ERATO~\cite{ERATO} were used;
the studies of the \lX\ and
\jjX\ final states used calculations based on the program DELTGC~\cite{DELTGC},
cross-checked with GRC4F~\cite{GRC4F}; 
DELTGC and NUNUGPV~\cite{NUNUGPV}
were used to calculate the signals expected in the \gX\ topology. The EXCALIBUR and
GRC4F models were interfaced to the JETSET hadronization model~\cite{JETSET} 
tuned to $Z$ data~\cite{dtune}. 
The study of backgrounds due to $q{\bar q}(\gamma)$
production was made using events from the PYTHIA
model~\cite{PYTHIA}, while EXCALIBUR was used to study the $q\overline{q}\nu\overline{\nu}$
contribution to the \jjX\ topology, and KORALZ~\cite{KORALZ},
BHWIDE~\cite{BHWIDE} and TEEG~\cite{TEEG} 
were used in the calculation of backgrounds in the \lX\
final state. PYTHIA and EXCALIBUR were used in the simulation of events from
$ZZ$ production. Two-photon backgrounds were studied using the generators of
Berends, Daverveldt and Kleiss~\cite{BDK} and the TWOGAM
generator~\cite{TWOGAM}. 
All of these generators were interfaced to the full DELPHI simulation
program~\cite{detector} except DELTGC and ERATO, which were used only
to calculate event weights as a function of the trilinear gauge coupling
parameter values (see section~\ref{sec:methods}).

\vspace{0.5cm}
\noindent {\bf Selection of events in the \jjlv\ topology:}
\vspace{0.2 cm}

\noindent Events in the \jjlv\ topology are characterised by two hadronic jets,
a lepton and missing momentum resulting from the neutrino. The lepton may be an
electron or muon (coming either from $W$ decay or from the cascade decay of
the $W$ 
through a $\tau$ lepton) or, in the case of $\tau$ decays, the $\tau$  might
give rise to a low multiplicity jet. 
The major backgrounds come from $q\bar{q}(\gamma)$
production and from four-fermion final states containing two quarks and two
leptons of the same flavour.

Events with several hadrons were selected by requiring 5 or more charged
particles and a total energy of charged particles recorded in the detector
exceeding 15\% of the centre-of-mass energy. In the selection of $jj\mu\nu$ and
$jje\nu$ events, the candidate lepton was assumed to be the most energetic
charged particle in the event; for $jj\tau\nu$ events, the lepton
candidates were constructed by looking for an isolated $e$ or $\mu$ or a low
multiplicity jet.

The selection procedure was identical to that used in our analysis of data at
183~GeV~\cite{DELPHI183}, except that, 
in the selection of electron candidates, the component of the missing momentum
transverse to the beam axis was required to be greater than 15~GeV/$c$ and the
angle between the electron candidate and the missing momentum was required to exceed
60\dgr.

The efficiency for the selection of \jjlv\ events was evaluated using fully
simulated events to be $(79.3 \pm 0.2$)\%, $(59.4 \pm 0.3)$\% and $(31.7 \pm
0.3)$\%  
for muon, electron and tau events,
respectively. Using data taken only when all essential components of the
detector were operational, corresponding to an integrated luminosity of
149~pb$^{-1}$, 263 muon, 212 electron and 146 tau candidate events were
selected. A background contamination of $(0.226 \pm 0.016)$~pb 
was estimated, of which 58\%
came from the $q \bar{q} (\gamma)$ final state, 22\% from $Z\ee$, 13\% from
$ZZ$ and $Z\gamma^*$ production and small contributions from non-semileptonic
$WW$ events and other sources. The errors on the efficiencies and background
contributions (where given) are statistical errors, resulting from the quantity
of simulated data available. 
The systematic uncertainty resulting from these statistical
errors is included in the results shown in section~\ref{sec:results}.

\vspace{0.5cm}
\noindent {\bf Selection of events in the \jjjj\ topology:}
\vspace{0.2cm}

\noindent The selection of events in the fully hadronic topology followed
closely that used in our analysis of data at 183~GeV~\cite{DELPHI183}, with only
small changes in the values of kinematic cuts. 

All detected particles were first clustered into jets using LUCLUS~\cite{JETSET} with
$d_{join}=5.5$~GeV/$c$. Events were accepted if they had at least four jets,
with at least four  particles per jet. Background from $Z (\gamma)$ events was
suppressed by imposing  the condition $\sqrt{s^{\prime}} >130$~GeV, where
$\sqrt{s^{\prime}}$ is an estimate 
of the effective collision energy in the (background) $q\bar{q}(\gamma)$ final
state after initial state radiation~\cite{sprime}. 
Events were
then forced into a 4-jet configuration and a 4-constraint fit was performed,
requiring conservation of four-momentum. Then, in order to suppress the dominant
background, which arises from the $q \bar{q} (\gamma)$ final state, the condition
$D>0.0055$~GeV$^{-1}$ was imposed, with $D = \frac{E_{min}}{E_{max}} 
\theta_{min}/(E_{max}-E_{min})$; $E_{min}$ and $E_{max}$ are the energies of the
reconstructed jets with minimum and maximum energy and $\theta_{min}$ is the
minimum interjet angle in radians. A further fit was then performed on surviving events,
imposing four-momentum conservation and requiring the masses of the two 
reconstructed $W$s to be equal. The fit was applied to all three possible
pairings of the four jets into two $W$s. Fits with reconstructed $W$ mass
outside the range  $74 < m_{W}^{rec}< 88$~GeV/$c^2$ were rejected and, of the
remaining fits, the one with minimum $\chi^{2}$ was accepted. 

The efficiency of the selection procedure was evaluated from fully simulated
events to be $(75.7 \pm 0.2)$\%. A total of 1130 events was selected from data
corresponding to an integrated luminosity of 154.4~pb$^{-1}$. Background
contributions of $(1.26 \pm 0.02)$~pb and $(0.187 \pm 0.007)$~pb were estimated
from $q \bar{q} (\gamma)$ and \jjlv\ production, respectively.  The method used
in the analysis of the data to assign the reconstructed jets to $W$ pairs was applied to a
sample of simulated events generated with PYTHIA, with only the three doubly
resonant CC03~\cite{CC03} 
diagrams for $WW$ production present in the production amplitude; in this model
the efficiency of the procedure was estimated to be about 74\%.

An additional problem in the analysis of the \jjjj\ state is to distinguish the 
pair of jets constituting the \Wp\ decay products from that from the \Wm. This
ambiguity can be partly resolved by computing jet charges from the
momentum-weighted charge of each particle belonging to the jet, $Q_{jet}= \sum_i
q_i |p|_i^{0.5} / \sum_i |p|_i^{0.5}$ (where $q_i$ and $p_i$ are the charge and
the momentum of the particle and the exponent is chosen 
empirically), 
and defining the $W^\pm$ charges, $Q_{W^+}$ and $Q_{W^-}$, as the
sums of the charges of the two daughter jets. Following the method
of~\cite{alephjjjj}, the distribution of the difference $\Delta Q =
Q_{W^-}-Q_{W^+}$ was then used to construct an
estimator $P_{W^-}(\Delta Q)$ of the probability that the pair with the more
negative value of $Q_W$ is a \Wm. 

An estimate of the efficiency of this procedure was made (for the same sample of
simulated $WW$ events as was used to estimate the jet pairing efficiency)
by flagging the jet pairs with $\Delta Q < 0$ as $W^-$ and comparing with the
generated information.   
In order to separate this estimate from that for the efficiency of $W$ pair assignment,
only events with correct jet pairing were included in the 
comparison, leading to a value of 77\% for the $W$ charge tagging efficiency.

\vspace{0.3cm}
\noindent {\bf Selection of events in the \lX\ topology:}
\vspace{0.2cm}

\noindent In the selection of candidates for the \lX\ final state, events  were
required to have only one charged particle, clearly identified as a muon from
the signals recorded in the barrel or forward muon chambers or as an electron
from the signals in the barrel or forward electromagnetic calorimeters. The
corresponding selection criteria are described in detail in~\cite{W97}. In
addition, the normal track selections were tightened: the track was required to
pass within 1~mm of the interaction point in the $xy$ plane (perpendicular to
the beam) and within 4~cm in $z$. Lepton candidates were also required to have
momentum below 75~GeV/$c$, with a component transverse to the beam above 20~GeV/$c$. Events
were rejected if there was an energy deposition of more than 5~GeV in the barrel
or forward electromagnetic calorimeters which was not associated with the
charged particle track, or if there was any signal in the hermeticity
detectors.  In the selection of electron candidates, the ratio of the energy
measured in the electromagnetic calorimeter to the magnitude of the measured
momentum was required to exceed 0.7.

Imposing these criteria, 10 events were selected in data in the $e
X$ channel, and an efficiency of $(31.2 \pm 3.8)$\% was obtained  for
$ee\nu\nu$ production.   
In the $\mu X$ channel, 11 events were selected with a calculated efficiency of $(51.2
\pm  6.3)$\% for $e \mu \nu \nu$ production.
In both cases, the efficiency was evaluated in the phase space region
defined by the following cuts: one (and only one) lepton was emitted at more than
10.3\dgr, with an energy between 20~GeV and 75~GeV.

For Standard Model values of the couplings,
$8.3 \pm 1.0$ single electron events were expected, comprising 4.5 events
from $ee\nu\nu$ production, 0.3 events from $e\mu\nu\nu$, 0.5 events from
$e\tau\nu\nu$ with the $\tau$ decay products unseen, 0.2 
 events from the same final state but with an electron or positron
from the $\tau$ decay observed in the detector, and
2.8 events from the reaction $\ee \ra \ee\gamma(\gamma)
$ with one electron (or positron) and the final state photon(s) unobserved.
In the single muon channel, $10.1 \pm 1.7$ events were expected for Standard Model values
of the couplings, comprising 4.2 events from $e \mu \nu \nu$
production, 2.4 events from $ee \mu \mu$ production (coming mainly
from two-photon interactions), 0.3
events from $e \tau \nu \nu$, 0.6 events from  $\mu \mu
\nu \nu$, 0.4 events from $\mu \tau \nu \nu$, 2.1 events from $\mu \mu
\gamma$, and a negligible contribution from $\tau \tau \gamma$
production.

All the contributing channels except the Bhabha
and Compton backgrounds in the $eX$ final state and the $\mu\mu\gamma$ and $\tau\tau\gamma$
backgrounds and two-photon interactions in the $\mu X$ channel have a
dependence on trilinear gauge boson 
couplings in their production, and this was taken into account in the subsequent
analysis.

\vspace{0.5cm}
\noindent {\bf Selection of events in the \jjX\ topology:}
\vspace{0.2cm}

\noindent Events were selected as candidates for the \jjX\ topology
if they had  total measured transverse momentum greater than
20~GeV/$c$ and invariant mass of detected particles between 45 and 
90~GeV/$c^2$. The detected particles were clustered into jets using LUCLUS with
$d_{join}=6.5$~GeV/$c$, and events were accepted if they had two or three
reconstructed jets. Surviving events were then forced into a 2-jet
configuration.

Events from the $WW$ final state with one $W$ decaying leptonically were
suppressed by rejecting events with identified final state leptons ($e$ or
$\mu$) of energy exceeding 12~GeV. In order to suppress the contribution from
the $q{\bar q}(\gamma)$ final state, events were rejected if the acoplanarity
was greater than 160\dgr, where acoplanarity is defined as the angle between
the projections of the jet momenta on to the plane perpendicular to the beam. 
In addition, events were rejected if the polar angle of either
reconstructed jet was below 20\dgr, 
if any charged or
neutral particle of momentum exceeding 1~GeV/$c$ was reconstructed within a cone of
angle 30\dgr\ about the direction of the missing momentum, or if there was a
signal in the hermeticity detectors in a cone of angle 50\dgr\ about the
direction of the missing momentum.

Applying these criteria to fully simulated events, an efficiency of $(43.7
\pm 1.5)$\% was calculated; in the data 64 events were selected.
As in the \lX\ topology,
the efficiency is quoted with respect to a reduced phase space: the electron
had to be emitted at less than 10.3\dgr, 
the angle between the missing momentum vector (calculated as the negative of
the vector sum of the $q$ and $\bar{q}$ momenta) and the beampipe had to be
larger than 25\dgr, the visible transverse momentum
had to be at least 15~GeV/$c$ and the visible energy had to be 
less than 160~GeV. Both momentum and energy here are taken from the
four-momenta of the final state $q$ and $ \bar{q}$.
For the $q \bar{q}$ pair, acollinearity and
acoplanarity had to be less than 170\dgr, and the polar angle of both $q$ and
$\bar{q}$ had to exceed 20\dgr.

For Standard Model values
of the couplings, a total of $60.3 \pm 0.8$ events are expected, comprising 17.0 
events from the $q {\bar q} e \nu$ final state with the electron or
positron lost in the beam pipe, 5.1 events from $q {\bar q} e \nu$
with the electron or positron elsewhere in the detector, 22.0 events
from $q {\bar q} \tau \nu$, 8.1 events from $q {\bar q} \mu \nu$, 
3.9 events from $q {\bar q} \nu \nu$, 4.0 events from $q
{\bar q} (\gamma)$ production, and 0.2 events from $\gamma \gamma$
interactions. All the processes contributing to the selected sample except
$q{\bar q}(\gamma)$ production and two-photon interactions include diagrams with
trilinear gauge couplings, and this was taken into account in the subsequent
analysis.

\vspace{0.5cm}
\noindent {\bf Selection of events in the \gX\ topology:}
\vspace{0.2cm}

\noindent 
The production of the single photon final state, \gX, via a $WW\gamma$ vertex
proceeds through the fusion diagram shown in figure~\ref{fig:Fdiags}c, while the
dominant process giving rise to this final state, $\ee \ra Z \gamma$ with $Z
\ra \nu {\bar \nu}$, involves bremsstrahlung diagrams. The sensitivity of the
\gX\ final states to anomalous $WW\gamma$ couplings is therefore greatest when
the photon is emitted at high polar angle. Events were selected if they had a
single shower in the barrel electromagnetic calorimeter with $45\dgr <
\theta_\gamma < 135\dgr$ and $E_\gamma > 6$~GeV, where $\theta_\gamma$ and
$E_\gamma$ are the polar angle and energy, respectively, of the reconstructed
photon. It was also required that no electromagnetic showers were present in the
forward electromagnetic calorimeters, and a second shower in the barrel
calorimeter was accepted only if it was within $20\dgr$ of the first one. 
Cosmic ray events were suppressed by requiring any signal in the hadronic
calorimeter to be in the same angular region as the signal in the electromagnetic
calorimeter and the electromagnetic shower to point towards the beam
collision point~\cite{Vansinggam}. 
Using these criteria, 145 events were
selected from data corresponding to an integrated luminosity of 155~pb$^{-1}$.  
The Standard Model expectation is $157.7 \pm 3.7$ events. 
Values for the triple gauge boson couplings were fitted in the region $E_\gamma
> 50$~GeV, which contained 59\% of the events. In this region, an overall
selection efficiency of $(54 \pm 4)$\%  was estimated~\cite{Vansinggam},
with  negligible background contamination.

\section{Methods used to determine the couplings}
\label{sec:methods}

The analysis procedures applied are similar to those used in our
previously reported analysis of data at 183~GeV~\cite{DELPHI183}, though somewhat different
applications of the method of Optimal Observables were used in the
analyses of the \jjlv\ and \jjjj\ final states.

\vspace{0.5cm}
\noindent {\bf Optimal Observable analysis of \jjlv\ and \jjjj\ channels}
\vspace{0.2cm}

Data in both the \jjlv\ and \jjjj\ channels were analyzed using methods
based on that of Optimal Observables~\cite{OO}. The methods exploit the fact
that the differential cross-section, $d\sigma/d\vec{V}$,  where $\vec{V}$
represents the phase space variables, is quadratic in the trilinear gauge
coupling parameters: 

\ba
\frac{d\sigma(\vec{V},\vec{\lambda})}{d\vec{V}} 
   = c_{0}(\vec{V}) + \sum_{i}c^{i}_{1}(\vec{V})\cdot\lambda_{i} 
    +  \sum_{i \leq j}c_{2}^{i j}(\vec{V})\cdot \lambda_{i}\cdot\lambda_{j} \, ,
\label{eq:OO}
\ea 

\noindent where the sums in $i,j$ are over the set  $\vec{\lambda} = \{
\lambda_{1}, \ldots ,\lambda_{n} \}$ of parameters under consideration.  It has
been shown that the ``Optimal Variables'' $c^{i}_{1}(\vec{V})/c_{0}(\vec{V})$
and $c^{i j}_{2}(\vec{V})/c_{0}(\vec{V})$, approximated for real data by using
the reconstructed phase space variables $\vec{\Omega}$ as arguments of the
$c^{i}_{1}$ and $c^{i j}_{2}$, have the same estimating efficiency as can be
obtained in unbinned likelihood fits of parameters $\lambda_i$ to the
data~\cite{MO}. 

In the determination of a single parameter $\lambda$, the joint distribution of
the quantities  $c_{1}(\vec{\Omega}) / c_{0}(\vec{\Omega})$ and
$c_{2}(\vec{\Omega}) / c_{0}(\vec{\Omega})$ was compared with the expected
distribution, computed from events generated with EXCALIBUR and passed through
JETSET and the full detector simulation. 
An extended maximum likelihood fit, combining the information coming
from the shape of the Optimal Variables and from the cross-section, has been
carried out. At each stage the simulated data, which had been generated at a
few values of the couplings, have been 
reweighted~\cite{reweight} to the required value of $\lambda$ using the
matrix element calculation of the ERATO generator~\cite{ERATO}.  
In the case of
events in the \jjlv\ topology, the binning in these two variables was made
using a multidimensional clustering technique, described in detail
in~\cite{cluster}. This is an economical binning method in which the $n_d$ real
data points are used as seeds to divide the phase space into an equal number of
multidimensional bins. Each simulated event is associated with the closest real
event, resulting in an equiprobable division of the space of the Optimal
Variables in which it is assumed that the best available knowledge of the
probability density function is that of the real data points themselves. 

The use of such a technique becomes of particular importance when simultaneous
fits to two coupling parameters are performed. The number of Optimal Variables
then increases to five: $c_1^1/c_0$, $c_1^2/c_0$, $c_2^{1 1}/c_0$, $c_2^{2
2}/c_0$ and $c_2^{1 2}/c_0$, and the use of equal sized bins in a space of this
number of dimensions is impractical. For events in the \jjlv\ topology, an
extended maximum likelihood fit was performed over the $n_d$ bins for each
pair of coupling parameters ($\lambda_1, \lambda_2$) using this method.

A somewhat different technique was used in 2-parameter fits to data in the
\jjjj\ topology. In this case, extended maximum likelihood fits were made to 
the binned joint distribution of only the first order terms $c_1^1/c_0$ and
$c_1^2/c_0$ in (\ref{eq:OO}), but an iterative procedure was used, at each
stage expanding the expression for the differential distribution of the phase
space variables $\vec{V}$ about the values $(\tilde{\lambda_1},
\tilde{\lambda_2})$ obtained in the previous iteration:

\ba
\frac{d\sigma(\vec{V}, \lambda_1, \lambda_2)}{d\vec{V}} 
   = c_0(\tilde{\lambda_1}, \tilde{\lambda_2}, \vec{V}) 
     + c^1_1(\tilde{\lambda_1}, \tilde{\lambda_2}, \vec{V}) 
                                            (\lambda_1 - \tilde{\lambda_1}) 
     + c^2_1(\tilde{\lambda_1}, \tilde{\lambda_2}, \vec{V}) 
                                            (\lambda_2 - \tilde{\lambda_2}) 
     + ... \ .
\ea

\begin{figure}[htb]
\centerline{\epsfig
{file=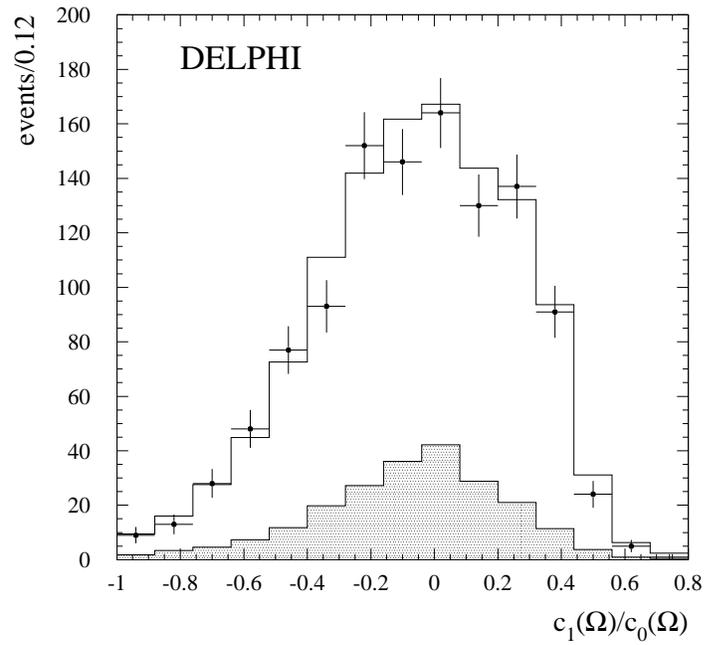,width=10cm} }
\caption{ 
Distribution of the optimal variable
$c_1^{\Delta g_1^Z}(\vec{\Omega}) / c_0(\vec{\Omega})$ (defined in the text) for the
coupling \Dgz\ in the \jjjj\ channel from DELPHI data at 189~GeV.  
The points represent the data and the histogram shows the distribution
expected for the fitted value of \Dgz\ (see table \ref{table:results1}). The shaded area is
the estimated background contribution.
}
\label{fig:jjjj_oo_dgz} 
\end{figure}

\noindent It has been shown in reference~\cite{MO} that when this iterative
procedure has converged sufficiently, the first order terms retain the whole
sensitivity of the Optimal Variables to the coupling parameters
$\vec{\lambda}$, the contribution from the higher order terms becoming
negligible. In practice, this was achieved after about three or four
iterations.
As an example, figure~\ref{fig:jjjj_oo_dgz} shows the distribution of $c_1^{\Delta
  g_1^Z}(\vec{\Omega}) / c_0(\vec{\Omega})$ for data and for the
results of the fit described in the next section.

\vspace{0.5cm}
\noindent {\bf Cross-check analysis of \jjlv\ and \jjjj\ channels}
\vspace{0.2cm}

In both the \jjlv\ and \jjjj\ channels, an additional analysis was performed
using more directly measured kinematic variables in order to corroborate results
obtained from the methods described above. 

\begin{figure}[t]
\centerline{\epsfig{file=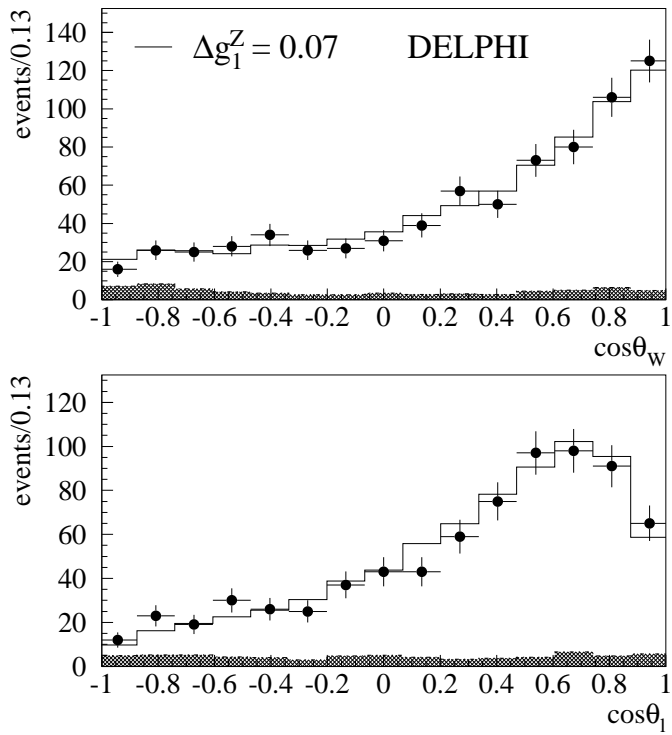, width=10cm}}
\caption
{Distributions in \cosw{} and \cosl{} for \jjlv\ events ($\ell \equiv
e,\mu, \tau$) for DELPHI data, shown as dots, at 189 GeV. 
The histogram shows the distribution expected for the fitted value of
\Dgz\ (see table \ref{table:results1}). 
The shaded area is the estimated background contribution. 
}
\label{fig:liv_wl}
\end{figure}

In the \jjlv\ topology, a
binned maximum likelihood fit was made to the joint distribution in \cosw, the
$W^-$ production angle, and \cosl, the polar angle of the produced lepton with
respect to the incoming $e^\pm$ of the same sign. In this study, somewhat looser
criteria were imposed in the selection of the events, giving a total sample of
743 semileptonic events, with estimated efficiencies of $(79.1 \pm 0.3)$\%,
$(67.3 \pm 0.4)$\% and $(40.4 \pm 0.5)$\% 
for muon, electron and tau events, respectively, and an estimated background
contamination of 0.49~pb. A 4-constraint kinematic fit was then applied to the
events, requiring conservation of four-momentum, and the variables \cosw\ and
\cosl\ computed from the fitted four-vectors. The expected number of events in
each bin was estimated using events generated with PYTHIA  corresponding to the
reaction \eeWW\ and passed through the full detector simulation procedure.
Again, a reweighting technique was used to determine the expected number of
events for given values of the coupling parameters.
The distributions of $\cos \theta_W$ and $\cos \theta_l$ for real and
simulated data are shown in figure~\ref{fig:liv_wl}.

In the \jjjj\ topology, the second analysis involved a binned extended maximum
likelihood fit to the production angular distribution. Events were selected by
constructing a probability function from the distributions of eleven kinematic
variables, namely: the value of $d_{join}$ in the LUCLUS algorithm when four
rather than three natural jets are reconstructed; the sphericity; the angle
between the two most energetic jets; the minimal multiplicity in a jet;
the second Fox-Wolfram moment; the $D$ variable (defined above); $s'$ (defined
above); the fitted  $W$ masses; the product of the energy ratios of the two jets
in the two reconstructed dijets; the minimal transverse momentum with respect to
the beam axis of the 15 most energetic particles in the event; the transverse
momentum  of the jet pair obtained by forcing the reconstruction of exactly two
jets. Using this procedure, a sample of 1331 events was selected with estimated
efficiency of $(86.6 \pm 0.2)$\% and purity of $(74.4 \pm 0.4)$\%. As in the
case of the optimal observable analysis of this channel described above,
momentum-weighted jet 
charges were then calculated to try to distinguish the \Wp\ decay products from
those of the \Wm. 
An angular variable 
$x_g = \cos\theta_W (P_{W^-}(\Delta Q) - P_{W^+}(\Delta Q))$,
was constructed from the cosine of the $W$ production angle and the
difference in probability of a dijet to come from a \Wm\ or \Wp\ decay. 
The experimental distribution of $x_g$ was
compared with predictions obtained from events generated with PYTHIA, passed
through the full detector simulation procedure, and reweighted in the fit for
given values of the coupling parameters.

\vspace{0.5cm}
\noindent {\bf Analysis of \lX, \jjX\ and \gX\ channels}
\vspace{0.2cm}

Data in the topologies \lX\ and \jjX\ were analysed using maximum likelihood
fits to the observed total numbers of events selected, while  the \gX\ data were
fitted using a binned extended maximum likelihood fit to the distribution of the
reconstructed photon energy, $E_\gamma$, in the region $E_\gamma > 50$~GeV,
which has the maximum sensitivity to anomalous triple gauge boson couplings.

\section{Results}
\label{sec:results}

\begin{figure}[t]
\centerline{\epsfig {file=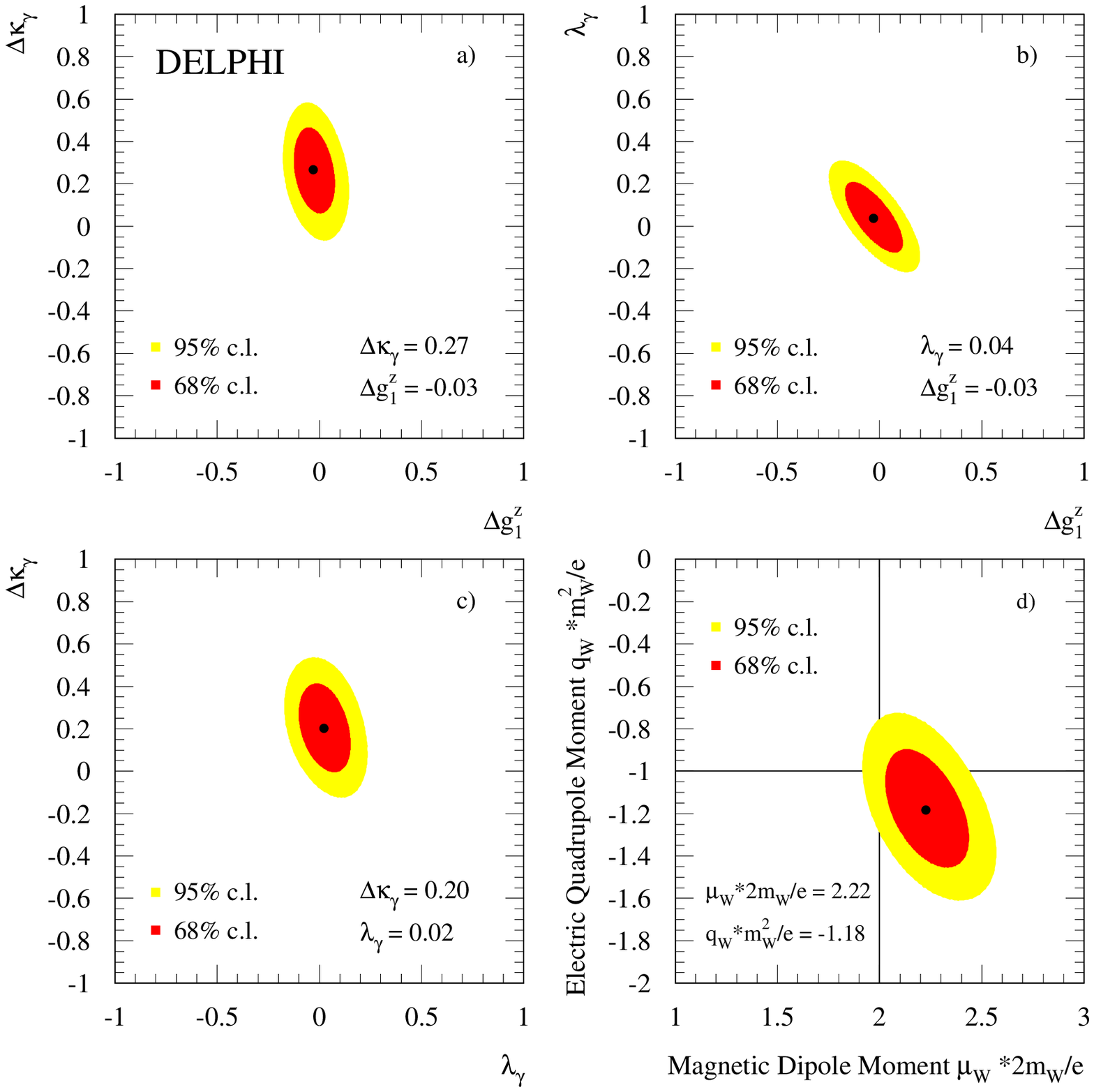,width=\textwidth} }
\caption[]
{Results of fits in the planes of the parameters a) (\Dgz,~\Dkg), b) (\Dgz,~\lm),
c) (\lm,~\Dkg) and d) ($\mu_W$,~$q_W$) using data from the final states listed in 
table~\protect\ref{table:results1} combined with DELPHI results at lower
energy~\protect\cite{DELPHI183}. In the combination, the analyses of the \jjlv\
and \jjjj\ final states based on Optimal Observable techniques were used. In
each case the third parameter was  fixed at its Standard Model value. The values
maximizing the likelihood function and the regions accepted at the 68\% and 95\%
confidence levels are shown. The confidence intervals are computed as the contours
where the value of the likelihood function is increased by 1.15 units (68\%
CL) and 3.0 units (95\% CL) respectively from the minimum.
}
\label{fig:twopar} 
\end{figure}

The results obtained for the triple gauge boson couplings from the data in each
of the final states and using the methods discussed above are shown in
table~\ref{table:results1}, together with their statistical and systematic
errors (see below).   
The results from all topologies are combined with
those previously analysed by DELPHI at 183~GeV and reported in
reference~\cite{DELPHI183} to give the values of the coupling parameters, their
errors and the 95\% confidence limits shown in
table~\ref{table:results2}. In the combination, which is done by adding the
individual log-likelihood functions, the results in the \jjlv\ and
\jjjj\ topologies from the methods based on Optimal Observables were used, as
these use all the available kinematic information and hence are expected to have
greater precision. 
In the fit to each coupling parameter, the values of the
other parameters were held at zero, their Standard Model values. The results of
fits in which two of the couplings \Dgz, \Dkg\ and \lm\
were allowed to vary are shown in
figure~\ref{fig:twopar}a-c. In no case is any deviation seen from the Standard
Model prediction of zero for the couplings determined.

The results shown in figure~\ref{fig:twopar}c can 
be transformed to
produce estimates for the magnetic dipole moment, $\mu_W$, and the electric
quadrupole moment, $q_W$, of the $W$ boson using the relations

\ba
      \mu_W & = &  \frac{e}{2 m_W} (g_1^\gamma + \kg + \lm) \quad {\text{and}}  \\
        q_W & = &  -\frac{e}{m_W^2} (\kg - \lm) \, .
\ea
\noindent The resulting two-parameter fit gives the values
\ba
        \mu_W \cdot \frac{2 m_W}{e} & = & 2.22^{+0.20}_{-0.19}  \, \quad
        {\text{and}} \nonumber \\ 
         q_W  \cdot \frac{m_W^2}{e} & = & -1.18^{+0.27}_{-0.26} \, \nonumber
\ea

\noindent with the confidence level contours shown in figure~\ref{fig:twopar}d.
In the derivation of the result for $\mu_W$, the value of
$g_1^\gamma$, the $WW\gamma$ charge coupling, has been assumed to be unity, as
required by electromagnetic gauge invariance. The quantity $(g-2)_W$, derived
from the definition of the gyromagnetic ratio of a particle of spin $\vec{s}$, charge
$Q$ and mass $m$, $\vec{\mu} = g\vec{s}\frac{Q}{2m}$, is, therefore, measured
to be $(g-2)_W = 0.22 ^{+0.20}_{-0.19}$.

\vspace{0.5cm}
\noindent {\bf Systematic uncertainties:}
\vspace{0.2cm}

The systematic errors shown in table~\ref{table:results1} and included in the
results shown in table~\ref{table:results2} 
contain contributions from various sources. 
Table~\ref{table:systematics} lists the dominant sources of systematic
uncertainties for each of the analyses used in the combination. 
A distinction between systematic errors affecting more than one channel and
effects specific to only one channel is made in the combination
of the different channels. The list of common systematic effects and the
procedure for their combination is given later in this section.

In the \jjlv{} channel, the dominant effect for \Dgz{} and \lm{} arises from
the uncertainty in the background contamination, where a conservative
estimate of $\pm 10\%$ was used. 
For \Dkg,
the event reconstruction effects give a comparable contribution. 
Comparisons between $Z$ data and fully simulated events were used to estimate
uncertainties of jet and lepton energies and of their angular distributions.
These uncertainties were then used to derive an additional smearing for
a sample of simulated events, which was then also fitted to the data.
The difference arising from fitting this sample and the standard sample is
quoted as the event reconstruction uncertainty.
A further effect considered was the possibility of misassignment of
the lepton charge. This was again studied in $Z$ data, where the fraction of
events with misidentified lepton charge was found to be 0.3\%.
The corresponding systematic effect was calculated by fitting to a 
simulated sample of \jjlv\ data with 0.3\% of the events randomly assigned
the wrong lepton charge. 
Also included in the table is the systematic error arising from effects of
limited Monte Carlo statistics in the evaluation of signal efficiencies.

In the \jjjj{} channel, significant contributions to the systematic error
come from the use of simulated event samples with energies different from
that of the data, conservatively evaluated by comparing samples generated at
188 and 190 GeV (and labelled ``beam energy'' in the table), and from
uncertainties in the jet hadronization model used. 
The latter were estimated 
by comparing data sets in which the JETSET and HERWIG~\cite{HERWIG}
fragmentation models were applied to a common set of generated events. 
The effects of colour reconnection following the SK1 model~\cite{SK1} were investigated
by performing a similar comparison between a sample with maximal reconnection
probability and the standard unconnected set of JETSET events.
As in the analysis of the \jjlv{} channel, uncertainties due to
the background contamination (taken to be $\pm 5\%$) and from limited
simulated signal statistics were also taken into account.

In the single $W$ channels \lX{} and \jjX{}, the dominant source of systematic
errors is the uncertainty in the efficiency estimation, which is an effect of the
limited amount of simulated events available. Limited statistics also affect
the background estimation.  

In the \gX\ channel, systematic effects play only a minor role. The main
systematic contribution originates from the  uncertainty in the energy
reconstruction of the barrel electromagnetic calorimeter.  

As in our previous analysis~\cite{DELPHI183}, the combined results shown in
table~\ref{table:results2} include the independent
systematic errors from each channel. 
In addition, systematic effects common to more than one channel, such as the theoretical
uncertainty in the $WW$ cross-section, and the uncertainties in the $W$ mass, in the
luminosity measurement and in the LEP beam energy were taken into account 
separately. 
The most interesting effect among these correlated systematics is the
uncertainty in the $WW$ cross-section calculation, labelled ``signal
cross-section'' in table ~\ref{table:systematics}. 
To estimate this effect, the cross-section was varied by its theoretical
error of $\pm$2\%.
The effect of this variation is quite small,
particularly in the \jjlv{} channel, which contains the highest sensitivity
to the couplings studied. 
This is reassuring given that a more precise evaluation of the cross-section
is now available~\cite{NewWWx-sec}, which gives a value around
2\% lower than currently assumed. As this new cross-section calculation is
not yet implemented in our event generators, we have used the old calculations and quote a
systematic error which covers the difference between the two cross-section
values.  
Uncertainties in the differential cross-sections that could arise
from the difference between our Monte Carlo generators and the new
generators, or from the theoretical uncertainty in the new
calculations are the subject of an ongoing LEP-wide study and are not taken
into account in the results presented here.

The common effects were evaluated individually for each final state and then
added with weights derived from the statistical precision of the individual
channels with respect to each coupling. 

\section{Conclusions}
\label{sec:Conclusions}

Values for the $WWV$ couplings \Dgz, \Dkg\ and \lm\  have been derived from an
analysis of DELPHI data at 189~GeV. The results have been combined with
previously published values from DELPHI data at 183~GeV, giving an
overall improvement in precision by a factor of about two over that of the
183~GeV data~\cite{DELPHI183}. The results of the 2-parameter fit to the
couplings \Dkg\ and \lm\ have been used to derive values for the magnetic dipole
and electric quadrupole moments of the $W$ and for the $W$ gyromagnetic ratio.

There is no evidence for deviations from Standard Model predictions in any of
the results obtained. 

\subsection*{Acknowledgements}
\vskip 3 mm

We would like to thank our technical collaborators, our funding agencies for
their support in building and operating the DELPHI experiment, and the CERN
SL divsion for the excellent performance of the LEP collider. \\
We acknowledge in particular the support of \\
Austrian Federal Ministry of Science and Traffics, GZ 616.364/2-III/2a/98, \\
FNRS--FWO, Flanders Institute to encourage scientific and technological 
research in the industry (IWT), Belgium,  \\
FINEP, CNPq, CAPES, FUJB and FAPERJ, Brazil, \\
Czech Ministry of Industry and Trade, GA CR 202/96/0450 and GA AVCR A1010521,\\
Commission of the European Communities (DG XII), \\
Direction des Sciences de la Mati$\grave{\mbox{\rm e}}$re, CEA, France, \\
Bundesministerium f$\ddot{\mbox{\rm u}}$r Bildung, Wissenschaft, Forschung 
und Technologie, Germany,\\
General Secretariat for Research and Technology, Greece, \\
National Science Foundation (NWO) and Foundation for Research on Matter (FOM),
The Netherlands, \\
Norwegian Research Council,  \\
State Committee for Scientific Research, Poland, 2P03B06015, 2P03B11116 and
SPUB/P03/DZ3/99, \\
JNICT--Junta Nacional de Investiga\c{c}\~{a}o Cient\'{\i}fica 
e Tecnol$\acute{\mbox{\rm o}}$gica, Portugal, \\
Vedecka grantova agentura MS SR, Slovakia, Nr. 95/5195/134, \\
Ministry of Science and Technology of the Republic of Slovenia, \\
CICYT, Spain, AEN96--1661 and AEN96-1681,  \\
The Swedish Natural Science Research Council,      \\
Particle Physics and Astronomy Research Council, UK, \\
Department of Energy, USA, DE--FG02--94ER40817.

\clearpage
\begin{table}[p]
\begin{center}
\small{
\begin{tabular}{|l|c|c|c|}                                               \hline
        &     \Dgz                   & \Dkg             &        \lm  \\ \hline
\jjlv\ (Optimal Variables)     &
   $~~0.00^{+0.08}_{-0.08} \pm 0.02$ &                                          
                    $~0.28^{+0.35}_{-0.28} \pm 0.10$    &                       
                                     $~~0.06^{+0.09}_{-0.09} \pm 0.02$ \\ \hline
\jjlv\ (\cosw, \cosl)          & 
   $~0.07^{+0.12}_{-0.11} \pm 0.03$  &                                          
                    $~0.00^{+0.43}_{-0.24} \pm 0.10$    &                       
                                     $~~0.06^{+0.11}_{-0.10} \pm 0.03$ \\ \hline
\jjjj\ (Optimal Variables)     &
   $-0.09^{+0.14}_{-0.12} \pm 0.07$  &                                          
                    $~0.12^{+0.54}_{-0.31} \pm 0.24$    &                       
                                     $~~0.01^{+0.17}_{-0.15} \pm 0.05$ \\ \hline
\jjjj\ (\cosw)                 &
   $-0.07^{+0.17}_{-0.13} \pm 0.06$  &                                          
                    $~0.06^{+0.57}_{-0.31} \pm 0.23$    &                       
                                      $-0.05^{+0.19}_{-0.15} \pm 0.06$ \\ \hline
\lX\                           &
  $-0.45^{+1.35}_{-0.38} \pm 0.21$  &                                          
                    $~0.23^{+0.27}_{-0.34} \pm 0.19$    &                       
                                     $~~0.48^{+0.33}_{-1.27} \pm 0.21$ \\ \hline
\jjX\                          &
 $ -0.43^{+1.31}_{-0.39} \pm 0.25$   &                                          
                    $~0.19^{+0.34}_{-0.57} \pm 0.11$    &                       
                                     $~~0.42^{+0.36}_{-1.20} \pm 0.15$ \\ \hline
\gX\                           &
             --                      &                                          
                    $~0.70^{+0.77}_{-0.99} \pm 0.03$    &                       
                                     $~~0.65^{+1.03}_{-1.79} \pm 0.09$ \\ \hline
\end{tabular}

}
\caption[]{
Fitted values of $WWV$ coupling parameters from DELPHI data at 189~GeV using
the methods described in the text. 
The first error given for each value is the statistical error at 68\%
confidence level (CL),  
obtained by stepping up 0.5 units from the minimum of the likelihood curve; 
the second is the systematic error. In the fits to each
parameter, the others were set to zero, their Standard Model values.
}
\label{table:results1}
\end{center}
\end{table}

\begin{table}[p]
\begin{center}
\small{
\begin{tabular}{|c|r|c|}                                                  \hline
Coupling parameter &    Value  \qquad \qquad & 95\% confidence interval \\ \hline
\Dgz  &      $~-0.02 ^{+0.07}_{-0.07} \pm {0.01} $  &$-0.16 , 0.13 $            \\ \hline
\Dkg  &      $~~0.25 ^{+0.21}_{-0.20} \pm {0.06} $  &$-0.13 , 0.68 $            \\ \hline
\lm   &      $~~0.05 ^{+0.09}_{-0.09} \pm {0.01} $  &$-0.11 , 0.23 $            \\ \hline
\end{tabular}
}
\caption[]{
Values of $WWV$ coupling parameters combining DELPHI data from various
topologies and energies, 
 as described in the text. The second column shows the value of each
parameter corresponding to the minimum of the combined negative log-likelihood
distribution and its errors at 68\% CL. 
The first error quoted is the combined statistical and uncorrelated   
systematic error, the second is the total common systematic (see text).
The third column shows the 95\% confidence
intervals on the parameter values, computed by stepping up 2.0 units from the
minimum of the likelihood curve. 
In the fits to each coupling parameter, the
other two parameters were set to zero, their Standard Model values.
}
\label{table:results2}
\end{center}
\end{table}

\begin{table}[p]
\begin{center}
\small{
\begin{tabular}{|cc|c|c|c|}                                                  \hline
Channel & Source \& Method & \Dgz  & \Dkg & \lm  \\ \hline
\jjlv & Background estimation &  $ \pm 0.013$   & $ \pm 0.058 $  & $ \pm 0.014 $  \\ \hline
 & Signal cross-section &  $ \pm 0.002 $   & $ \pm 0.018 $  & $ \pm 0.002 $ \\ \hline
 & Lepton charge assignment &  $ \pm 0.005$   & $ \pm 0.035 $  & $ \pm 0.009 $ \\ \hline
 & Signal MC statistics  &  $ \pm 0.005$   & $ \pm 0.017 $  & $ \pm 0.006 $ \\ \hline 
 & Event reconstruction &  $ \pm 0.005$   & $ \pm 0.064 $  & $ \pm 0.006 $ \\ \hline
 & Total \jjlv{} systematic &  $ \pm 0.017$   & $ \pm 0.097 $  & $ \pm 0.019 $  \\ \hline
 \hline
\jjjj & Background estimation &  $ \pm 0.02 $   & $ \pm 0.05 $  & $ \pm 0.01 $ \\ \hline
 & Signal cross-section &  $ \pm 0.02 $   & $ \pm 0.13 $  & $ \pm 0.01 $ \\ \hline
 & Colour reconnection &  $ \pm 0.03 $   & $ \pm 0.07 $  & $ \pm 0.01 $ \\ \hline
 & Fragmentation &  $ \pm 0.01 $   & $ \pm 0.11 $  & $ \pm 0.03 $ \\ \hline
 & Beam energy &  $ \pm 0.05 $   & $ \pm 0.11 $  & $ \pm 0.02 $ \\ \hline
 & Total \jjjj{} systematic &  $ \pm 0.07$   & $ \pm 0.24 $  & $ \pm 0.05$  \\ \hline 
\hline
\lX & Background estimation &  $ \pm 0.13$   & $ \pm 0.13 $  & $ \pm 0.12 $  \\ \hline  
 & Signal cross-section &  $ \pm 0.08$   & $ \pm 0.05 $  & $ \pm 0.07 $  \\ \hline
 & Efficiency estimation &  $ \pm 0.15 $   & $ \pm 0.13 $  & $ \pm 0.16 $ \\ \hline
 & Total \lX{} systematic &  $ \pm 0.21$   & $ \pm 0.19 $  & $ \pm 0.21 $  \\ \hline 
\hline
\jjX  & Background estimation &  $ \pm 0.03$   & $ \pm 0.02 $  & $ \pm 0.03 $  \\ \hline
 & Signal cross-section &  $ \pm 0.08$   & $ \pm 0.06 $  & $ \pm 0.08 $  \\ \hline
 & Efficiency estimation &  $ \pm 0.23 $   & $ \pm 0.09 $  & $ \pm 0.12 $ \\ \hline
 & Total \jjX{} systematic &  $ \pm 0.25$   & $ \pm 0.11 $  & $ \pm 0.15 $  \\ \hline
\hline
\gX & Energy reconstruction &  $-$   & $ \pm 0.03 $  & $ \pm 0.09 $ \\
\hline
 & Signal cross-section &  $-$   & $ \pm 0.01 $  & $ \pm 0.01 $ \\ \hline
 & Efficiency estimation &  $-$   & $ \pm 0.01 $  & $ \pm 0.01 $ \\ \hline
 & Total \gX{} systematic &  $-$   & $ \pm 0.03 $  & $ \pm 0.09$  \\ \hline 
\end{tabular}
}
\caption[]{
Main systematic contributions in each analysed channel.
}
\label{table:systematics}
\end{center}
\end{table}

\clearpage

\end{document}